\documentclass[conference]{IEEEtran}

\usepackage[bottom = 1.1in, left=0.65in, top=0.75in,right=0.65in ]{geometry}
\usepackage{amsmath,amsfonts}
\usepackage{amsthm}
\usepackage{algorithmic}
\usepackage{algorithm}
\usepackage{array}
\usepackage[caption=false,font=normalsize,labelfont=sf,textfont=sf]{subfig}
\usepackage{textcomp}
\usepackage{stfloats}
\usepackage{verbatim}
\usepackage{graphicx}
\usepackage{cite}
\usepackage{color}
\graphicspath{{Figure/}}

\usepackage{url}
\newtheorem{prop}{Proposition}
\newtheorem{theo}{Theorem}

\begin{document}

\title{Towards Green Communication: Soft Decoding Scheme for OOK Signals in Zero-Energy Devices}

\author{\IEEEauthorblockN{
Ticao Zhang, 
Dennis Hui, 
Mehrnaz Afshang,  and  
Mohammad Mozaffari 
\vspace{0.2cm}
}

\IEEEauthorblockA{\small Ericsson Research, Silicon Valley, Santa Clara, CA, USA. \\ Emails: \{ticao.zhang, dennis.hui,  mehrnaz.afshang, mohammad.mozaffari\}@ericsson.com}
}


\maketitle

\begin{abstract}

The booming of Internet-of-Things (IoT) is expected to provide more intelligent and reliable communication services for higher network coverage, massive connectivity, and low-cost solutions for 6G services. However, frequent charging and battery replacement of these massive IoT devices brings a series of challenges.
Zero energy devices, which rely on energy-harvesting technologies and can operate without battery replacement or charging, play a pivotal role in facilitating the massive use of IoT devices. In order to enable reliable communications of such low-power devices, Manchester-coded on-off keying (OOK) modulation and non-coherent detections are attractive techniques due to their energy efficiency, robustness in noisy environments, and simplicity in receiver design. Moreover, to extend their communication range, employing channel coding along with enhanced detection schemes is crucial. In this paper,  a novel soft-decision decoder is designed for OOK-based low-power receivers to enhance their detection performance.  In addition, exact closed-form expressions and two simplified approximations are derived for the log-likelihood ratio (LLR), an essential metric for soft decoding. Numerical results demonstrate the significant coverage gain achieved through soft decoding for convolutional code.

\end{abstract}


\section{Introduction}
Zero-energy devices, requiring no maintenance, batteries, or manual charging, have the capability to harness energy from the world around them~\cite{ericsson_report}. These devices hold the potential to take on various roles, functioning as sensors to collect data, trackers to monitor locations, or actuators to prompt other machinery.  Many of these devices are deployed in harsh environments, such as agriculture fields, marine or embedded in bodies, where frequent battery replacement is not feasible. 
Therefore, minimizing device power consumption is crucial for zero-energy devices as it enables efficient utilization of energy harvesting, ensuring sustained operation while reducing dependence on external energy sources. To realize the full potential of these devices, {\color{black}low-complexity, power-efficient receiver designs that are resilient in noisy environments} have received wide attention recently~\cite{naser2023zero,zhang2023toward,guruacharya2020optimal,devineni2021manchester}. 

In industry and standardization organizations, there is significant attention to the design of low-power devices, with a focus on the following two primary applications~\cite{ericsson2021zeroenergy,Oppo,3GPP2021}. The first involves the enablement of low-power devices as ``wake-up radios", acting in conjunction with the primary radio to minimize its power consumption~\cite{3GPP_TR_38869_2023}. Currently, discussions on wake{\color{black}-up} radio are ongoing within the 3rd Generation Partnership Project (3GPP) Release 18 study on wake-up receivers and wake-up signals \cite{3GPP_TR_38869_2023}. The second direction encompasses low-power devices serving as zero-energy or ambient Internet of Things (IoT) devices {\color{black} that are} reliant on energy harvesting~\cite{3gppTR38848}. Such ambient IoT devices are currently being investigated within 3GPP Release 19 and IEEE ambient power (AMP)~\cite{ieee80211amp}.

For designing low-power devices, OOK modulation with simple envelope detection stands as one of the most popular choices due to its simplicity, lower circuit complexity and reduced energy consumption~\cite{zhang2011investigations,devineni2020non,guruacharya2020optimal
}. Furthermore, the adoption of Manchester-coded OOK introduces additional benefits, encompassing clock recovery, improved noise immunity, DC balance, and self-synchronization \cite{tang2012wake,devineni2021manchester,zargari2023improved}. Another key aspect of supporting OOK waveforms is the possibility of generating OOK-like signals using  orthogonal frequency division multiplexing (OFDM) transmitters ensures compatibility with the existing OFDM-based technologies including the 5G New Radio (NR) ~\cite{zhang2023toward}.

Despite the simplicity of the OOK modulation, its waveforms are prone to channel fadings and may incur a high bit error rate (BER). Thus, there is a need for {\color{black} enhanced detection} and coding schemes for OOK signals that are received by low-power receivers. To extend transmission range, enable error detection and correction, and enhance data reliability, channel coding is essential for these low-power devices.

\subsection{Related Work}
The performance of noncoherent detection of OOK signals has been extensively studied in the literature. The capacity behavior of noncoherent OOK over additive white Gaussian noise (AWGN) channels in the low signal-to-noise ratio (SNR) regime was investigated in \cite{zhang2011investigations} for wireless sensor networks (WSNs) from the point of information theory and the authors show that both the soft decision and hard decision of the noncoherent OOK modulation can achieve the Shannon limit.  In \cite{guruacharya2020optimal}, the optimal non-coherent OOK signal detector for ambient backscattering communication system was studied. Two types of detectors were designed based on the joint probability density function (pdf) of the received OOK signals. The noncoherent multi-antenna receiver design for direct OOK modulation for an ambient backscatter system was studied in \cite{devineni2020non}. To further improve the SNR gain in the ambient backscattering system, \cite{devineni2021manchester} explored to use of Manchester encoding with OOK signals, and the proposed scheme was shown to achieve a substantial performance improvement.
More recently, {\color{black}an improved detection scheme based on the correlation of the received OOK-modulated signals was proposed} to address the problem of channel uncertainties in ambient communication systems {\color{black} \cite{zargari2023improved}}. 

{\color{black}Computing Log-Likelihood Ratios (LLR) is crucial in the receiver performance as they provide an optimal soft decision information for the channel decoder.} The LLR values for OOK modulation under AWGN channel conditions are provided in~\cite{nagaraj2013soft}. 
However, to our knowledge, the evaluation of LLRs for Manchester-coded OOK modulation with a simple envelope detector under fading channel conditions {\color{black} with unknown noise power} has not been addressed in the existing literature.

\subsection{Our Contributions}
In this paper, we investigate the signal detection performance of the OOK signals with convolutional code for forward error protection and Manchester encoding.  Specifically, we designed a soft-decision decoder for the low-power receivers to enhance their detection performance and improve the reliability of the communication. For soft-decision OOK, analytical expressions and its low-complexity approximations for the LLR are derived, based on which {\color{black} a simple non-coherent detector with nearly optimal performance is obtained}. Simulation results show that the proposed non-coherent OOK
modulation with a soft-decision detection scheme can significantly improve the performance of the traditional hard-decision detection scheme, especially with proper interleaving in block Rayleigh fading channels.


\section{System Model and LLR computation}

\subsection{System Model}
\begin{figure*}[!t]
\centering
\includegraphics[width=0.6\textwidth]{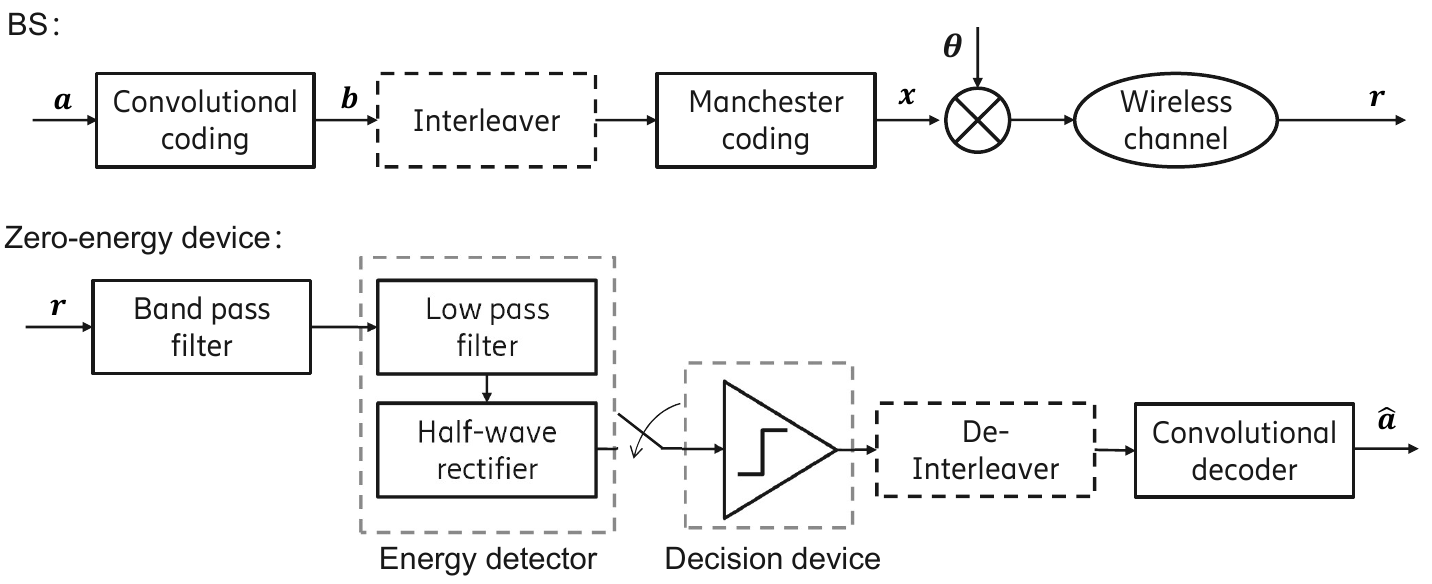}
\caption{Illustration of the proposed soft OOK signal transmission system including BS and low-power receiver.}
\label{fig:model}
\end{figure*}

Consider a downlink system where the base station (BS) transmits the OOK signals to the zero-energy devices as shown in Figure \ref{fig:model}. 
At the transmitter side (i.e., BS), the signal is Manchester encoded with convolutional code. At the receiver side, a simple energy detector and decision module are employed to recover the transmitted signals. 

To be specific, the desired transmitted 
 bit sequence has a length of $L$ and can be denoted as $\mathbf{a}\in\mathbb{R}^{L}$. To ensure reliable communication across a wide range of channels and environments, convolutional code $g(\cdot)$ that maps the bit sequence to the encoded signal is used, and it is expressed as $\mathbf{b}=g(\mathbf{a})\in \mathbb{R}^{M}$ where $M-L$ redundant bits are added for error protection. Moreover, to allow for easier clock synchronization between the transmitter and receiver, Manchester encoding is used where the transition from 1 to 0 represents
bit ``1” and the transition from 0 to 1 represents bit ``0”. Now we denote $\bar{\mathbf{x}}\in \mathbb{R}^{2M}$ as the Manchester encoded sequences of $\mathbf{b}$.
For the ease of signal  analysis in the time domain, we 
 further fill each bit of $\bar{\mathbf{x}}$ with $T$ repeated samples and get the time domain samples as $\mathbf{x}\in\mathbb{R}^{2MT}$. Note that  $\mathbf{x}$ can be split into $M$ Manchester bit periods and each period has $T$ samples for the first half of the Manchester-coded signal and $T$ samples for the second half of the Manchester-coded signal. We use $x_i(t)$ to represent the $t$-th sample within the $i$-th Manchester bit period ($1\le i\le M$, $1\le t\le 2T$) in $\mathbf{x}$.

Considering that the low-power receiver employs envelope detection on incoming signals, which exclusively extracts information from the signal's amplitude. The low-power receiver becomes insensitive to the phase of the transmitted samples. The additional degree of freedom in the phase of the generated baseband signal can be harnessed in several ways. First, it can be utilized to create an OOK-like signal from the OFDM signal, ensuring compatibility with the existing NR structure~\cite{zhang2023toward}. Second, extra data, such as additional coded bits or cyclic redundancy checks, can be modulated onto the phase for more capable receivers. Third, the phase can be adjusted to meet requirements for power spectral density (PSD) at the transmitter. In this paper, without loss of any generality,  we consider a case where a random phase sequence is modulated to the OOK signals, i.e., the transmitted signal is modeled as $x_i(t)e^{j\theta_i[t]}$ where $\theta_i[t]$ is the modulated phase.

As the time domain signal goes through a channel, the received signal can be modeled as $r_i(t)=h_ix_i(t)e^{j\theta_i(t)}+n_i(t)$ where $h_i$ is the block fading channel, 
  $n_i(t)\sim\mathcal{CN}(0,\sigma^2)$ is the channel noise and $\sigma^2$ is the noise power. We assume that $\theta_i(t)$ is uniformly distributed. Now we are interested in decoding the target binary sequence $\mathbf{b}$ from the amplitude of the received signals $\mathbf{r}$. 

\subsection{LLR Computation}
The log-Likelihood Ratio (LLR), defined in (\ref{eq:LLR}) is a pivotal metric that quantifies the reliability of the received data bits and improves error correction in noisy channels. Soft-decision decoding relies on LLRs, enhancing performance by considering both `'0" and `'1" likelihoods. LLRs play a crucial role in iterative decoding for advanced error-correcting codes. They capture channel characteristics and noise, optimizing decoding algorithms. In addition, LLRs are vital in adaptive modulation, selecting the right scheme for varying channel conditions.
\begin{equation} \label{eq:LLR}
{\rm LLR}=
\log\frac{P(b=1|
{\text{received signal})}}{P(b=0|\text{received signal})}.
\end{equation}
 Now, we start by evaluating $P(b=1|{\text{received signal}})$. At the receiver, for the $i$th bit period ($i=1,2,...,M$) at time slot $t$, the probability of transmitted bit is ``1" is
\small
\begin{align}
  & \quad P(b_i=1|\{|r_i(t)|\}_{t=1,2,...,2T},h_i) \nonumber \\
         = & \quad \frac{P(b_i=1)}{P(\{|r_i(t)|\}_{t=1,2,...,2T},h_i)}P(\{|r_i(t)|\}_{t=1,2,...,2T}|b_i=1,h_i)  
         \nonumber
\end{align}
\normalsize
Similarly, the probability of transmitted bit is ``0" and can be written as
\small
\begin{align}
  & \quad P(b_i=0|\{|r_i(t)|\}_{t=1,2,...,2T},h_i) \nonumber \\
         = & \quad \frac{P(b_i=0)}{P(\{|r_i(t)|\}_{t=1,2,...,2T},h_i)}P(\{|r_i(t)|\}_{t=1,2,...,2T}|b_i=0,h_i). \nonumber
\end{align}
\normalsize
Here, we exploit the Manchester encoding to provide a clock signal embedded within the data itself for the low-power receiver. The transition in the middle of each bit (from high to low or low to high) serves as a clock edge. This self-clocking feature simplifies clock recovery at the low-power receiver to maintain synchronization between the transmitter and receiver. The transition from `one' to
`zero' represents bit “1”. Hence, the probability of the received signal conditioned on $b_i =1$ and $h_i$ is 
\begin{align}
    & P(\{|r_i(t)|\}_{t=1,2,...,2T}|b_i =1,h_i)  \nonumber\\
     = & \prod_{t=1}^{T}P(|r_i(t)||\bar{x}_{2i-1}=1,h_i) \cdot \prod_{t=T+1}^{2T} P(|r_i(t)||\bar{x}_{2i}=0,h_i). \nonumber
\end{align}
Similarly, the transition from `zero' to `one' represents bit “0” and the probability of the received signal conditioned on $b_i =0$ and block fading $h_i$ is 
\begin{align}
    & P(\{|r_i(t)|\}_{t=1,2,...,2T}|b_i=0,h_i)  \nonumber\\
     = & \prod_{t=1}^{T}P(|r_i(t)||\bar{x}_{2i-1}=0,h_i) \prod_{t=T+1}^{2T} P(|r_i(t)||\bar{x}_{2i}=1,h_i). \nonumber
\end{align}

At the receiver, we could model the received signal as
$|r_i(t)|e^{j\phi_i(t)}$ where $\phi_i(t)$ is the phase that has a distribution $f(\phi)$. Note that at the transmitter, a uniformly distributed phase is modulated to the amplitude of the OOK signals. Hence, $\phi_i(t)$ is also uniformly distributed. We use $f(\phi)$ and $f(\theta)$ to denote the uniform distributions of $\phi$ and $\theta$, respectively. Then the pdf of the amplitude of the received signal conditioned on $b_i=1$ can be expressed as
\small
\begin{align}\label{eq:p1}
    & P(|r_i(t)||b_i=1,h_i) \nonumber\\
    = &   \int_0^{2\pi}\int_0^{2\pi}  \frac{1}{\pi\sigma^2} \exp\left( -\frac{||r_i(t)|e^{j\phi}-h_ie^{j\theta}|^2}{\sigma^2}  \right)f(\phi)f(\theta){\rm d}\phi{\rm d}\theta \nonumber\\
  =  &  \frac{1}{\pi\sigma^2} \exp\left( -\frac{|r_i(t)|^2+|h_i|^2}{\sigma^2}  \right) \int \int g(\theta,\phi)f(\phi)f(\theta) {\rm d}\phi {\rm d}\theta,  
\end{align}
\normalsize
where $g(\theta,\phi)=\exp\left( \frac{2|r_i(t)||h_i|\cos(\phi-\theta-\alpha)}{\sigma^2}\right)$ and $\alpha$ is the angle of $h_i$.

Similarly, the probability of the transmitted bit is ``0" can be expressed as
\small
\begin{align}\label{eq:p0}
    P(|r_i(t)||b_i=0,h_i) 
    & = 
    \int_0^{2\pi} \frac{1}{\pi\sigma^2} \exp\left( -\frac{||r_i(t)|e^{j\phi}|^2}{\sigma^2}  \right)f(\phi){\rm d}\phi \nonumber\\
     & = \frac{1}{\pi\sigma^2} \exp\left( -\frac{|r_i(t)|^2}{\sigma^2}  \right).
\end{align}
\normalsize
Having these probabilities a closed-form analytical expression for LLR is provided in the next Theorem.

\begin{theo}\label{Thm_LLR_exact}[LLR-exact] The exact closed-form expression of LLR for Manchester encoded OOK signal for the non-coherent receiver can be expressed as:
\small
\begin{align}\label{eq:LLR_exact}
    {\rm LLR}_i = \sum_{t=1}^T\log I_0(\frac{2|r_i(t)||h_i|}{\sigma^2}) - \sum_{t=T+1}^{2T}\log I_0(\frac{2|r_i(t)||h_i|}{\sigma^2}), 
\end{align}
\normalsize
where $I_0(x)=\frac{1}{\pi}\int_0^{\pi}e^{x\cos(z)}{\rm d}z$ is the modified zero-order Bessel function.
\end{theo} 

{\textit{Proof:}}
When the prior probability of the transmitted bit is equal, i.e., $P(b_i=0)=P(b_i=1)=0.5$, the LLR for the $i$th bit can be computed as
\small
\begin{align}\label{eq:LLR_ex}
    &  {\rm LLR}_i =  \log\frac{P(b_i=1|\{|r_i(t)|\}_{t=1,2,...,2T},h_i)}{P(b_i=0|\{|r_i(t)|\}_{t=1,2,...,2T},h_i)} \nonumber\\
   =&   \log \frac{P(\{|r_i(t)|\}_{t=1,2,...,2T}|b_i=1,h_i)}{P(\{|r_i(t)|\}_{t=1,2,...,2T}|b_i=0,h_i)} \nonumber\\
   =&  \log\frac{\prod\limits_{t=1}\limits^{T}P(|r_i(t)||\bar{x}_{2i-1}=1,h_i) \prod\limits_{t=T+1}\limits^{2T} P(|r_i(t)||\bar{x}_{2i}=0,h_i)}{\prod\limits_{t=1}\limits^{T}P(|r_i(t)||\bar{x}_{2i-1}=0,h_i) \prod\limits_{t=T+1}\limits^{2T} P(|r_i(t)||\bar{x}_{2i}=1,h_i)} \nonumber\\
   =&  \sum_{t=1}^T \log\int_0^{2\pi} \int_0^{2\pi}g(\theta,\phi)f(\phi)f(\theta) {\rm d}\phi {\rm d}\theta \nonumber\\
   -&  \sum_{t=T+1}^{2T} \log\int_0^{2\pi} \int_0^{2\pi}g(\theta,\phi)f(\phi)f(\theta) {\rm d}\phi {\rm d}\theta.
\end{align}
\normalsize

To compute the double integral in (\ref{eq:LLR_ex}), we use a substitution for the inner integral. Let $u=\theta+\alpha-\phi$, then the inner integral becomes
\small
\begin{align}
    \int_{\alpha-\phi}^{2\pi+\alpha-\phi}\exp\left( \frac{2|r_i(t)||h_i|\cos(u)}{\sigma^2}    \right){\rm d}u =2\pi I_0(\frac{2|r_i(t)||h_i|}{\sigma^2}). \nonumber
\end{align}
\normalsize

\subsection{Low-complexity approximations to the LLR}
\begin{figure}[!t]
\centering
\includegraphics[width=0.38\textwidth]{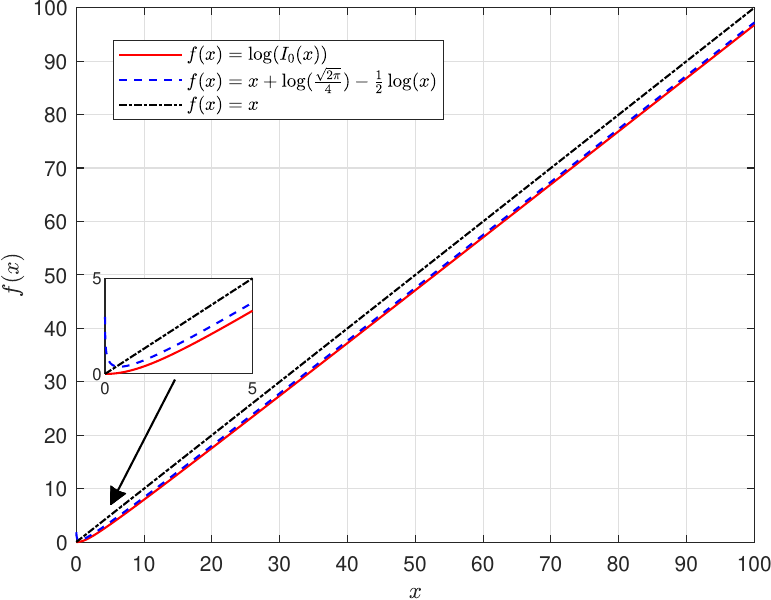}
\caption{Illustration of function $f(x)=\log(I_0(x))$ and its low-complexity approximations.}
\label{fig:appro}
\end{figure}

Low-power receiver devices have limited computational processing capability and requirements for power consumption optimization. Therefore, it is always desirable to streamline and simplify the LLR computation. In order to simplify the LLR expression presented in Theorem~\ref{Thm_LLR_exact}, we begin by simplifying $\log(I_0(x))$. 
{\color{black} Note that $\cos(z)=1-2\sin(z/2)^2$, then we have 
\begin{align}
    I_0(x) & = \frac{1}{\pi}\int_0^{\pi} e^{x\cos(z)}{\rm d}z = e^x\underbrace{ \frac{1}{\pi} \int_0^{\pi} e^{-2x\sin(z/2)^2} {\rm d}z }_{p(x)}
\end{align}

On the interval $[0,\pi]$, $\sin(z/2)\ge z/\pi$. As a result, we have
\begin{align}
    p(x) \le \frac{1}{\pi} \int_0^{\pi}e^{-2x(\frac{z}{\pi})^2} {\rm d}z
 < \frac{1}{\pi} \int_0^{\infty}e^{-2x(\frac{z}{\pi})^2}       {\rm d}z 
 = \frac{1}{4}\sqrt{\frac{2\pi}{x}} \nonumber
\end{align}
and
\begin{align}\label{eq:appr_O}
    \log(I_0(x))& \le \log\left(\frac{1}{4}e^x\sqrt{\frac{2\pi}{x}}\right) \nonumber\\
    & = x+\log\left(\frac{\sqrt{2\pi}}{4}\right)-\frac{1}{2}\log(x) \nonumber \\
    & =x\left(1+\mathcal{O}\left(\frac{\log(x)}{x}\right)\right)
\end{align}
 when $x$ is large, the value of $\log(x)$ is relatively small compared with $x$. Therefore, the term $\mathcal{O}\left(\frac{\log(x)}{x}\right)$ in (\ref{eq:appr_O}) can be dropped and we have $\log(I_0(x))\approx x$. 
 
 Figure~\ref{fig:appro} shows the approximations of the $f(x)=\log(I_0(x))$ function. It can be seen that negligible difference in approximation error can be achieved with the $f(x)=x+\log\left(\frac{\sqrt{2\pi}}{4}\right)-\frac{1}{2}\log(x)$ function especially when $x$ is larger than 10. There are some small differences in approximation error when using $f(x)=x$ function though, this is still a relatively good approximation. With this approximation, we therefore have the following proposition.
 \begin{prop}
The LLR value in (\ref{eq:LLR_exact}) can be approximated as
\begin{align}\label{eq:LLR_p1}
    {\rm LLR}_i & \approx  \frac{2|h_i|}{\sigma^2}\left(\sum_{t=1}^T |r_i(t)| - \sum_{t=T+1}^{2T}|r_i(t)|\right)  
\end{align}
\end{prop}

Note that (\ref{eq:LLR_p1}) involves the value of the channel $h_i$ and the noise variance $\sigma^2$. In practical scenarios, channel and noise estimation pose significant challenges for low-power receiver devices due to their limited computational capabilities and power constraints. Therefore, it is highly desirable to evaluate LLR without the need for accurate estimation of channel and noise variance. Considering that in block fading channels, the value of $h_i$ and $\sigma^2$ stay the same for each fading block and we can view $\frac{2|h_i|}{\sigma^2}$ as a scaling factor and we have the following proposition.

\begin{prop}
    For block fading channel, the LLR can be further simplified as
    \begin{align}
    {\rm LLR}_i \approx \sum_{t=1}^T |r_i(t)| - \sum_{t=T+1}^{2T}|r_i(t)|.
    \end{align}
\end{prop}
In this way, there is no need to perform further channel estimation at the receiver side, and only the amplitude of the received signal matters. In many low-power receiver device deployment scenarios, such as short-range wireless communication or stationary applications, we often encounter short-duration communication periods with relatively stable channel conditions which makes the block fading assumption quite reasonable for modeling and analysis. In the next section, we investigate the tightness of these approximations and compare the performance with hard-decision decoding, where the LLR value is computed as
\begin{equation}
{\rm LLR}_i =
\left\{
    \begin{array}{lr}
       1, \quad \text{if } \sum_{t=1}^T |r_i(t)|\ge \sum_{t=T+1}^{2T}|r_i(t)|,\\
        -1, \quad  \text{otherwise. } 
    \end{array}
\right.
\end{equation}

}


\section{Simulation Results and Discussions}
In this section, numerical results are presented to evaluate the performance of the proposed detection schemes.  For the block Rayleigh fading channel, we set $h_i$ to follow $\mathcal{CN}(0,1)$ while for AWGN $h_i=1$. The OOK bit sequences are convolutional coded with Manchester encoding. We use a two-memory convolutional encoder which has a polynomial generator of [15,13].  The bit sequence length is set as $L=1000$ and the samples per bit is  $T=2$.  All the channels hold unchanged for the same channel block and the channel block length is 1003. We investigate the  performance of the soft decision and hard decision decoding performance for the OOK signals through  Monte-Carlo simulations. 


\subsection{Performance in AWGN Channel}
\begin{figure}[!t]
\centering
\includegraphics[width=0.38\textwidth]{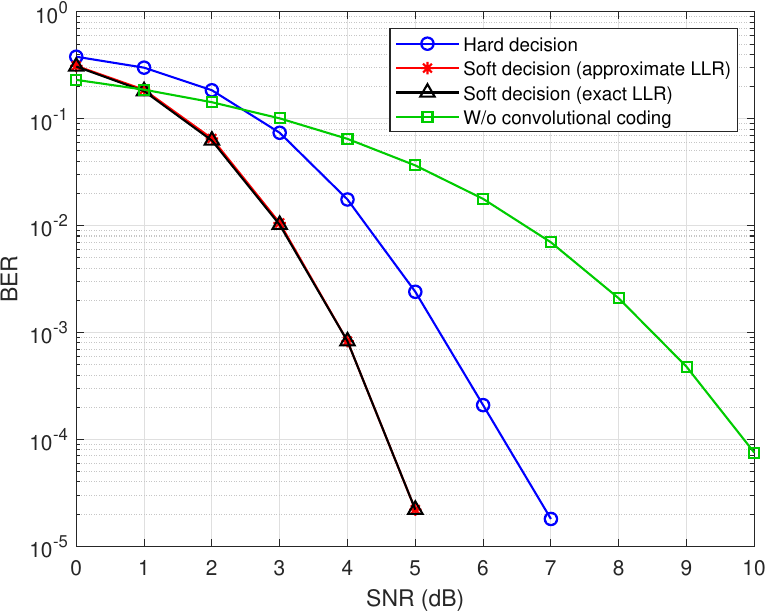}
\caption{BER performance of hard- and soft-decision decoding with convolutional codes in AWGN channels.}
\label{fig:BER_AWGN}
\end{figure}

Figure~\ref{fig:BER_AWGN} compares the BER performance for different decoding schemes in the AWGN channel. The proposed soft-decision detection schemes with exact LLR values achieve the best detection performance. As the channel knowledge and the noise variance value are difficult to obtain in practice, we also use the approximate LLR value to simplify the process. The soft decision scheme with approximate LLR  achieves a performance that is tightly close to that with the exact LLR value. This experiment suggests that in practice there is no need to perform channel and noise variance estimation at the low-power receiver side which further reduces the complexity and power consumption. Simply using the soft decision detection with approximate LLR value guarantees a satisfactory performance. Moreover, compared with hard decision detection, soft decision detection achieves around 1.5 dB gain at 0.1\% BER  in AWGN channels. Compared with direct transmission without the use of convolutional coding, the soft-decision detection achieves around 4.6 dB performance improvement.

\begin{figure}[!t]
\centering
\includegraphics[width=0.38\textwidth]{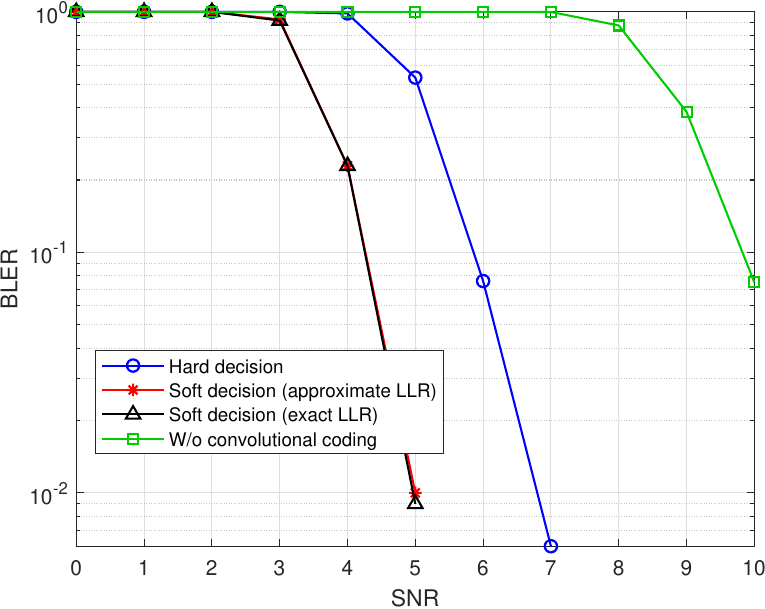}
\caption{BLER performance of hard- and soft-decision decoding with convolutional codes in AWGN channels.}
\label{fig:BLER_AWGN}
\end{figure}

The block error rate (BLER) for different detection schemes is shown in Figure~\ref{fig:BLER_AWGN}. As can be seen, the proposed soft-decision detection scheme outperforms the conventional hard-decision detection scheme by around 1.8 dB and outperforms the scheme without the use of convolutional code by around 5.6 dB at 10\% BLER. Again, soft-decision detection with approximate LLR value achieves similar performance to the soft-decision detection scheme with the exact LLR. 

\begin{figure}[!t]
\centering
\includegraphics[width=0.38\textwidth]{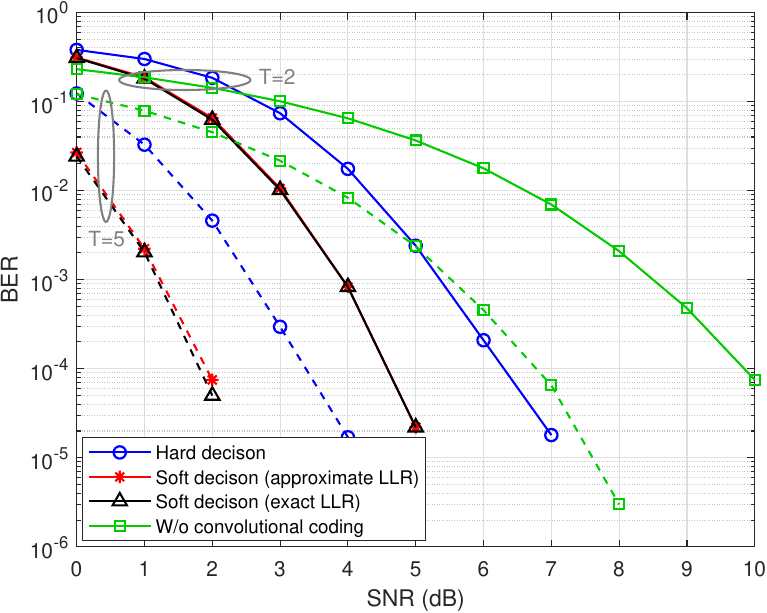}
\caption{The impact of time domain modulation in AWGN channels.}
\label{fig:BER_T}
\end{figure}

The impact of the number of time domain samples on the BER performance in AWGN channels is shown in Figure \ref{fig:BER_T}. As we increase the number of samples per bit from $T=2$ to $T=5$, the performance can be improved. This can be explained since the decoding schemes can be improved when more samples are used to represent one-bit information. 

\subsection{Performance in Block Rayleigh Fading Channel}
\begin{figure}[!t]
\centering
\includegraphics[width=0.38\textwidth]{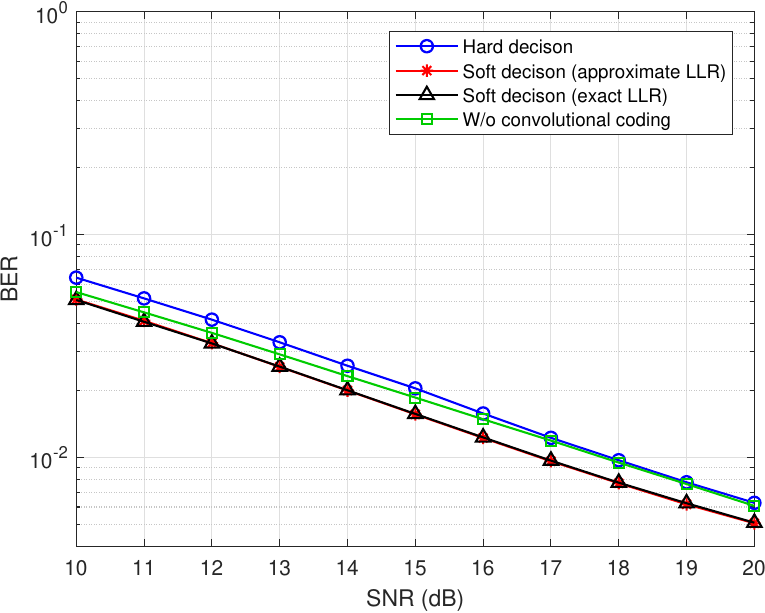}
\caption{BER performance of hard- and soft-decision decoding with convolutional codes in block Rayleigh fading channels.}
\label{fig:BER_Rayleigh}
\end{figure}

Figure \ref{fig:BER_Rayleigh} shows the BER performance of the decoding schemes with convolutional codes in Rayleigh fading channels. The proposed soft-decision decoding scheme outperforms the hard-decision decoding scheme by around 1 dB in block Rayleigh fading channels. However, compared with the performance in the AWGN channel as shown in Figure \ref{fig:BER_AWGN}, the BER performance does not improve significantly with the increase of the SNR. To combat this problem, we considered adding an interleaver at the transmitter and the low-power  receiver performs deinterleaving. 

\begin{figure}[!t]
\centering
\includegraphics[width=0.38\textwidth]{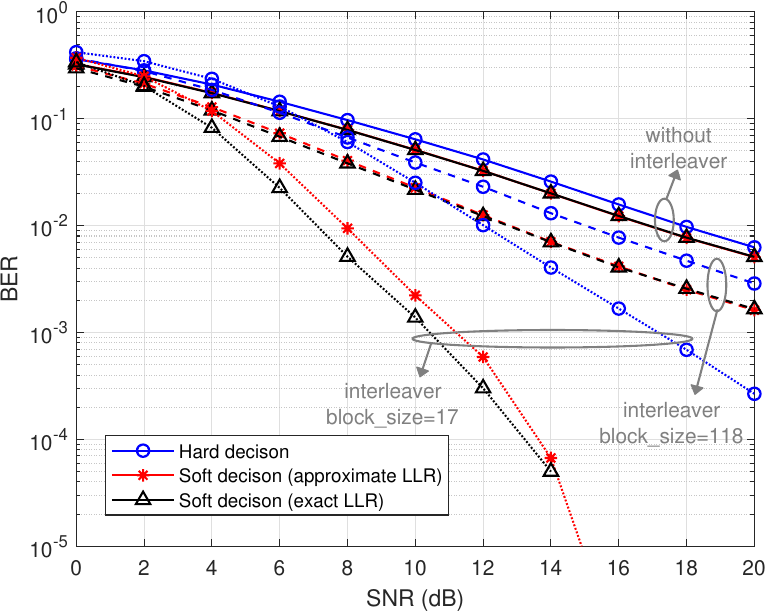}
\caption{The impact of the interleaver on the decoding performance in block Rayleigh fading channels.}
\label{fig:BER_Rayleigh_interleaver}
\end{figure}

The impact of interleaver on the performance is shown in Figure \ref{fig:BER_Rayleigh_interleaver}. We observe that by decreasing the interleaver block size, the performance of the soft decision scheme improves significantly. For example, at 1\% BER, the soft decision detection scheme with an interleaving block size of 118 improves the decoding performance by around 4.2 dB compared with the case without interleaving. By reducing the interleaving block size to 17 (or increasing the number of interleaving blocks), the performance can even improved by around 10 dB. This performance gain is significant, which suggests that proper interleaver along with the decoding schemes could together be used to combat channels fadings for OOK signals. Moreover, when the interleaver block size is 17, the proposed soft decision detection scheme with an exact LLR value outperforms the soft decision detection scheme with an approximate LLR value. 
Finally, when comparing the performance of hard-decision decoding with soft-decision decoding, we find that the introduction of interleaver brings more gains to the soft decision decoding scheme compared with the hard decision decoding scheme. This results further demonstrates the advantages of the use of interleaver and  proper soft decision decoding.

\subsection{Impact of the Randomly Modulated Phase Sequences}
\begin{figure}[!t]
\centering
\includegraphics[width=0.38\textwidth]{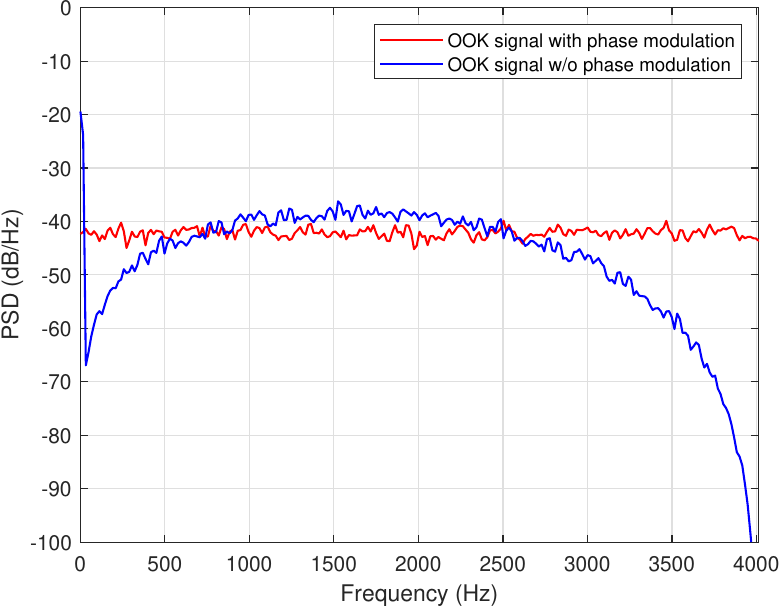}
\caption{The PSD of the convolutional encoded OOK signals.}
\label{fig:PSD}
\end{figure}

Note that our proposed soft-decision decoding scheme applies to not only simple OOK bit sequences but also phase-modulated OOK bit sequences. The main implementation concern is that when transmitting an ideal OOK sequence, the power spectrum density (PSD) is not flat and not favorable at the transmitter side as shown in Figure \ref{fig:PSD}. By modulating the OOK signal with a random phase, the PSD becomes favorable from an implementation perspective. Moreover, the introduction of the extra phase does not affect the amplitude of the OOK signals, which brings no impact on the noncoherent decoding at the receiver. Further, the extra modulated phase can be potentially used to convey extra information for possible coverage extension and error protection.   

\section{Conclusions}
In this paper, we investigated the decoding performance for the noncoherent OOK signals with convolutional
code error protection and Manchester encoding. We derived the LLR expressions and their low complexity approximations. Simulation results show that the proposed soft decision decoding scheme and its low complexity approximations achieve considerable gains in AWGN channels. Moreover, substantial performance gain can be achieved when the proposed soft decoding schemes are used together with the proper interleaving scheme in block Rayleigh fading channels. 
The LLR approximation does not require precise channel estimation and can avoid complex numerical computations. Hence, the proposed scheme has great potential in OOK signal decoding for low-power device communications.  

\bibliographystyle{IEEEtran}
\bibliography{main.bib}

\end{document}